\documentclass[prc,showpacs,showkeys]{revtex4}
\usepackage{epsfig,graphicx,float,color}
\usepackage{amssymb}
\usepackage[normalem]{ulem}

\begin{document}

\title{Phase diagram of the two-fluid Lipkin model: a butterfly catastrophe}
\author{J.E.~Garc\'{\i}a-Ramos$^1$, P.~P\'erez-Fern\'andez$^2$, J.M~Arias$^3$, E.~Freire$^4$}  
\affiliation{
  $^1$Departamento de  F\'{\i}sica Aplicada, Universidad de Huelva, 21071 Huelva, Spain\\
  $^{2}$Departamento de  F\'{\i}sica Aplicada III, Escuela T\'ecnica Superior de Ingenier\'{\i}a, Universidad de Sevilla, Sevilla, Spain\\
  $^{3}$Departamento de F\'{\i}sica At\'omica, Molecular y Nuclear, Facultad de F\'{\i}sica, Universidad de Sevilla, Apartado~1065, 41080 Sevilla, Spain \\
  $^{4}$Departamento de  Matem\'atica Aplicada II, Escuela T\'ecnica Superior de Ingenier\'{\i}a, Universidad de Sevilla, Sevilla, Spain\\}
\begin{abstract} 
\begin{description}
\item [Background:] In the last few decades quantum phase transitions have
  been of great interest in Nuclear Physics. In this context,
  two-fluid algebraic models are ideal systems to study how 
  the concept of quantum phase transition evolves when moving into more
  complex systems, but the number of publications along this line has
  been scarce up to now. 

\item [Purpose:] We intend to determine the phase diagram of a two-fluid
  Lipkin model, that resembles the nuclear proton-neutron interacting boson
  model Hamiltonian, using both numerical results and analytic tools,
  i.e., catastrophe theory, and to compare the mean-field results with exact
  diagonalizations for large systems.

\item [Method:] The mean-field energy surface of a
  consistent-Q-like two-fluid Lipkin Hamiltonian is studied and compared with
  exact results 
  coming from a direct diagonalization. The mean-field
  results are analyzed using the framework of catastrophe theory.   

\item [Results:] The phase diagram of the model is obtained and
  the order of the different phase-transition lines and surfaces is
  determined using 
  a catastrophe theory analysis. 
 
\item [Conclusions:] There are two first order surfaces
  in the phase diagram, one
  separating the spherical and the deformed shapes, while the other
  separates two different deformed phases. A
  second order line, where the later surfaces merge, is found. This line finishes in a
  transition point with a divergence in the second order derivative of
  the energy that corresponds to a tricritical point in the language
  of the Ginzburg-Landau theory for phase transitions.
\end{description}
\end{abstract}


\maketitle

\section{Introduction}
\label{sec-intro}
The study of quantum phase transitions (QPTs) is a hot topic in
different areas of quantum many-body physics. In Nuclear Physics many
aspects of QPTs have been studied \cite{Cast07,Cejn09,Cejn10}, both
theoretically and experimentally.  Also in other fields such as Molecular Physics \cite{Iach08,Pere11}, Quantum Optics \cite{Emar,Amic08} or Solid State Physics \cite{Sach11} studies related to relevant QPTs have been recently presented.

QPTs are phase transitions analogous to the classical ones but occurring at zero temperature. QPTs appear when the Hamiltonian has two (or more) parts with different structures (symmetries) and there is one (or several) control parameter that drives the system from one structure to the other. The phase transition is characterized by an abrupt change in an observable (called order parameter) that is zero in one phase and different from zero in the other. The value of the control parameter for which the structural change appears is known as critical value. Schematically a Hamiltonian undergoing a QPT is written as 
\begin{equation}
H(\xi)=\xi\cdot H(\mbox{symmetry}_1)+ (1-\xi)\cdot H(\mbox{symmetry}_2).
\end{equation}
For a particular value of the control parameter, $\xi_c$, which is the
critical value, the system undergoes a structural QPT from symmetry 1 to symmetry 2.

One interesting extension of the QPT concept appears when treating
with composed systems, as in the case of lattice systems \cite
{Iach15}. The simplest case is a quantum two-fluid system in which
there are two kind of particles (bosons in the case presented here)
with creation (and annihilation) operators that commute among
them. Some pioneering studies on two-fluid systems
\cite{Ar04,CI04,Caprio05} were carried out for the proton-neutron
interacting boson model, IBM-2, and the authors managed to construct
the phase diagram for a restricted Hamiltonian and classified the
different phase transitions that the system undergoes. Other models
that can be considered as two-fluid systems are the Dicke
\cite{Dick54} and the Jaynes-Cumming \cite{Jayn63} models for which
the two fluids correspond to photons and atoms. Note that in this case
the role of both fluids is not symmetric, photons fulfill a $hw(1)$
(Heisenberg-Weyl) algebra while atoms are governed by a $su(2)$ algebra. In the
case of IBM-2, both fluids are connected with a $u(6)$ algebra. 

The aim of this work is to study a simple two-fluid Lipkin model,
which corresponds to a $su(2)\otimes su(2)$ algebra. One of the main
motivations for carrying out this study is to treat with a model
somehow similar to IBM-2 (except for the $\gamma-$degree of freedom)
but simpler. In Refs.~\cite{Ar04,CI04}, when  discussing QPTs in IBM-2,
because of the large dimensions involved, exact results (obtained from
a direct diagonalization) were only obtained for small values of the
boson number. Thus, a comparison with the corresponding mean-field
results, valid for $N \rightarrow \infty$, was not
possible. Therefore, it is of interest to carry out such a comparison
for a model with similar physics content than IBM-2. In particular, it
has been shown that the IBM-1 and the Lipkin energy surfaces are
equal \cite{Vida06} and then, their phase diagrams are fully
equivalent. The advantage of the two-fluid Lipkin model with respect
to IBM-2 is the smaller dimensions involved, which allows one to
perform exact calculations with much larger boson
numbers. Consequently, this study will allow us to establish a proper
comparison with the mean-field results. Finally, it is worth noting
that Dicke and Jaynes-Cumming models correspond to a given limit of
the two-fluid Lipkin model, in which a {\it contraction} from $u(2)$
to $hw(1)$ is performed \cite{Iach06}.

The paper is organized as follows: in Section \ref{sec-model} the
algebraic structure of the model is outlined, while in Section
\ref{sec-cqf} the particular case of the consistent-Q like Hamiltonian
is worked out. Section \ref{sec-CL} is devoted to study the classical
limit of the model (mean field).  In Section \ref{sec-numerical} a
numerical study of the phase diagram is presented. Section
\ref{sec-taylor} is devoted to the application  of the catastrophe
theory in the study of the phase diagram and the unambiguous
determination of the order of the different phase
transitions. Finally, Section \ref{sec-conclusions} stands for the
summary and the conclusions.

\section{The Lipkin model and its two-fluid extension}
\label{sec-model}

The Lipkin model was proposed by Lipkin, Meshkov, and Glick in
\cite{Lipk65} as a toy model that is exactly solvable through a simple
diagonalization and appropriated to check the validity and limitations
of different approximation methods used in many-body physics (in
particular in Nuclear Physics). Since then, a plethora of applications
have appeared in the literature. Using a boson representation, the
model is built in terms of scalar bosons that can occupy two
non-degenerated energy levels labeled by $s$ and $t$. In the simplest
case, the building blocks are the creation $s^\dag$, $t^\dag$, and
annihilation $s$, $t$, boson operators. The four possible bilinear
products of one creation and one annihilation boson generate the
$u(2)$ algebra. If one combines two coupled Lipkin structures, one  obtains the two-fluid Lipkin model. In this model, there are two boson families identified by a subindex, $s_1^\dag$,  $t_1^\dag$ and  $s_2^\dag$, $t_2^\dag$, and the corresponding dynamical algebra will be $u_1(2)\otimes u_2(2)$,
whose generators are: $s^\dag_i s_i$, $s^\dag_i t_i$, $t^\dag_i s_i$, and $t^\dag_i t_i$, for $i=1,2$. If the boson number in each fluid is conserved, it is also of interest to consider the dynamical subalgebra $su_1(2) \otimes su_2(2)$ with generators
\begin{equation}
J^+_i=t^\dag_i s_i,\,~~~~~~~~ J^-_i=s^\dag_i t_i,\, ~~~~~~~~
J^0_i=\frac{1}{2}(t^\dag_i t_i-s^\dag_i s_i),
\end{equation} 
which verify the angular momentum commutation relations,
\begin{equation}
[J^+_i,J^-_i]=2J^0_i,\,~~~~~~~~~ [J^0_i,J^\pm_i]=\pm J^\pm_i.
\end{equation}
Adding the operators $N_i=s^\dag_i s_i + t^\dag_i t_i$ one recovers
the $u_1(2) \otimes u_2(2)$ algebra.

We consider that $s$ and $t$ present a different behavior with respect to the parity operator,      
\begin{eqnarray}
&~&P\, t^\dag_i\, P^{-1}= - t^\dag_i,\,  P\, t_i\, P^{-1}= -t_i,\nonumber\\
  &~&P\, s^\dag_i\, P^{-1}= s^\dag_i,\,  P\, s_i\, P^{-1}= s_i,
  \label{parity}
\end{eqnarray}
therefore, $s$ bosons preserve the parity while $t$ bosons do not,
i.e., $s$ has positive parity while $t$ has negative one. 
In $u(2)$ this assignment is arbitrary, but in higher-dimensional
models appears from physical considerations.

A detailed description of the $u_1(2) \otimes u_2(2)$ algebraic structure can be found in \cite{Fran94}. Here, we will present just an abridged version of that analysis. Starting from the dynamical algebra $u_1(2) \otimes u_2(2)$, the possible subalgebras chains are four: two of them correspond to an early coupling of the dynamical algebras into the direct-sum subalgebra $u_{12}(2)$ (or $su_{12}(2)$),
\begin{equation}
\begin{array}{cccccc}
u_1(2)\otimes u_2(2)&\supset&u_{12}(2)&\supset& u_{12}(1) & \\
\downarrow& &\downarrow& &\downarrow\\
N_1 \otimes N_2& &[h,h']& &n_t & \rightarrow \textrm{basis} ~~ |N_1~  N_2~ h~ n_t \rangle \\
\end{array},
\end{equation}
where the labels of the irreps verify the following branching rules: $h+h'=N_1+N_2$, $h \ge h'$, $1/2(N_1+N_2+h'-h)\le n_t\le
1/2(N_1+N_2-h'+h)$, and
\begin{equation}
\begin{array}{cccccccc}
u_1(2)\otimes u_2(2)&\supset&su_1(2)\otimes su_2(2)&\supset&su_{12}(2)&\supset& so_{12}(2)& \\
\downarrow& &\downarrow& &\downarrow& &\downarrow\\
N_1 \otimes N_2& &j_1 \otimes j_2& &j& &\mu & \rightarrow  \textrm{basis}~~ |j_1~  j_2~ j~ \mu \rangle \\
\end{array},
\end{equation}
where $j_i=N_i/2$,  $j=1/2(N_1+N_2),1/2(N_1+N_2)-1, ...,
1/2|N_1-N_2|$, $-j\le \mu\le j$, and $j=1/2(h-h')$. $|...>$ stands for
the basis state in the corresponding dynamical symmetry.

The second two algebras correspond to a late coupling into a direct-sum subalgebra,
\begin{equation}
\begin{array}{cccccc}
u_1(2)\otimes u_2(2)&\supset&u_1(1)\otimes u_2(1)&\supset&u_{12}(1)&\\
\downarrow& &\downarrow& &\downarrow\\
N_1 \otimes N_2& &n_{t_1} \otimes n_{t_2}& &n_{t}= n_{t_1}+ n_{t_2}  & \rightarrow \textrm{basis}~~  |N_1~  N_2~ n_{t_1}~ n_{t_2} \rangle \\
\end{array},
\end{equation}
where $n_{t_i} \le N_i$, and 
\begin{equation}
\begin{array}{cccccccc}
u_1(2)\otimes u_2(2)&\supset&su_1(2)\otimes su_2(2)&\supset& so_1(2)\otimes so_2(2)&\supset&so_{12}(2)& \\
\downarrow& &\downarrow& &\downarrow& &\downarrow\\
N_1 \otimes N_2& &j_1 \otimes j_2& &\mu_1 \otimes \mu_2& &\mu=\mu_1+\mu_2 &  \rightarrow  \textrm{basis}~~ |j_1 ~ j_2~ \mu_1~ \mu_2 \rangle \\
\end{array},
\end{equation}
where $-j_i\le \mu_i\le j_i$.

Concerning the Hamiltonian, the most general up to two-body interaction Hamiltonian can be written as,
\begin{equation}
H=H_1+H_2+H_{12},
\label{H1}
\end{equation}
where
\begin{eqnarray}
H_i&=&a_i {J'}_i^{0}+b_i(J^+ + J^-)+c_i (J_i^+ J_i^-)+d_i
((J_i^+)^2+(J_i^-)^2)+e_i(J_i^+ {J'}_i^{0}+ {J'}_i^{0} J_i^-)+f_i
({J'}_i^{0})^2\label{H2}\\
H_{12}&=&w_1(J_1^+ J_2^+ + J_1^- J_2^-)+w_2 (J_1^+ J_2^- + J_1^- J_2^+)+
w_3 (J_1^+ {J'}_2^{0}+J_1^- {J'}_2^{0})\nonumber\\
&+&
w_4({J'}_1^{0} J_2^+ + {J'}_1^{0}J_2^-)+ w_5 {J'}_1^{0} {J'}_2^{0},\label{H3}
\end{eqnarray}
being
\begin{equation}
{J'}_i^{0}={J}_i^{0}+\frac{N_i}{2}.  
\end{equation}
In an equivalent way the Hamiltonian can be expressed as,
\begin{eqnarray}
H_i&=&a_i t_i^\dag t_i + b_i(t_i^\dag s_i + s_i^\dag t_i )+
c_i (t_i^\dag s_i s_i^\dag t_i)+
d_i(t_i^\dag s_i t_i^\dag s_i+ s_i^\dag t_i s_i^\dag t_i)+
e_i(t_i^\dag s_i t_i^\dag t_i+ t_i^\dag t_i s_i^\dag t_i)+
f_i  t_i^\dag t_i t_i^\dag t_i
\label{Hi}\\
H_{12}&=&w_1(t_1^\dag s_1 t_2^\dag s_2 + s_1^\dag t_1 s_2^\dag t_2 )+
w_2 (t_1^\dag s_1 s_2^\dag t_2 + s_1^\dag t_1 t_2^\dag s_2)+
w_3 (t_1^\dag s_1 t_2^\dag t_2 + s_1^\dag t_1 t_2^\dag t_2)\nonumber\\
&+&
w_4( t_1^\dag t_1 t_2^\dag s_2+ t_1^\dag t_1 s_2^\dag t_2)+  
w_5 t_2^\dag t_2 t_1^\dag t_1 .
\label{H12}
\end{eqnarray} 

\subsection{The Consistent-Q-like Hamiltonian}
\label{sec-cqf}

A more restricted Hamiltonian, which is inspired in the consistent-Q-formalism of the IBM \cite{Ca}, will be used along the rest of the paper. This Hamiltonian resembles the schematic one used in many IBM-2 calculations, it was studied in detail in \cite{Ar04,Garc14} and will be the reference Hamiltonian in this work. The Hamiltonian can be written as,
\begin{equation}
H= x \left(n_{t_1}+n_{t_2} \right)-\frac{1-x}{N_1+N_2}
Q^{(y_1,y_2 )}\cdot Q^{(y_1,y_2 )}    
\label{Hcqf}
\end{equation} 
where
\begin{eqnarray}
n_{t_i}&=&t_{i }^{\dagger }t_{i },\\ 
Q^{(y_1,y_2)}&=&\left(Q_1^{y_1}+
Q_2^{y_2}\right), \\ 
Q_i^{y_i}&=& 
s_{i }^{\dagger } t_{i } + t_{i }^{\dagger }s_{i }+ 
y_i \left( t_{i}^{\dagger }{t}_{i}\right).
\end{eqnarray}
Due to the behavior of the bosons under parity (\ref{parity}), the Hamiltonian (\ref{Hcqf}) is in general non-parity conserving, except for $y_1=y_2=0$. This Hamiltonian (\ref{Hcqf}) can be obtained from the general one (\ref{H1},\ref{Hi},\ref{H12}) with the following relations among parameters,
\begin{eqnarray}
&~& a_i=x-2\frac{x-1}{N_1+N_2},\qquad b_i=\frac{x-1}{N_1+N_2}y_i,
\qquad c_i=2\frac{x-1}{N_1+N_2},\qquad d_i=\frac{x-1}{N_1+N_2}\\
&~& e_i=2 y_i \frac{x-1}{N_1+N_2},\qquad f_i=y_i^2\frac{x-1}{N_1+N_2},\qquad
\Delta_i=x-1,\\ 
&~&w_1=w_2=2\frac{x-1}{N_1+N_2},\qquad w_3=2 y_1\frac{x-1}{N_1+N_2},\qquad
w_4=2 y_2\frac{x-1}{N_1+N_2},\qquad w_5=2 y_1 y_2 \frac{x-1}{N_1+N_2}, 
\end{eqnarray}
(please note that $\Delta_i$ correspond to a shift in the energy origins) leading to the compact form, 
\begin{eqnarray}
H&=& x ({J'}_1^{0}+{J'}_2^{0})
-\frac{1-x}{N_1+N_2} 
(J_1^+ + J_1^-+J_2^+ + J_2^-+y_1 J_1'^0+y_2 J_2'^0)
(J_1^+ + J_1^-+J_2^+ + J_2^-+y_1 J_1'^0+y_2 J_2'^0)
\end{eqnarray}
This Hamiltonian is a mixture of dynamical symmetries of the problem,
particularly $u_{1}(1) \otimes u_{2}(1)$ for $x=1$, and $su_{1}(2)
\otimes su_{2}(2)$ for $x=0$ and $y_1=y_2=0$. 
This form is specially suitable to study QPTs, because one can associate a \textit{symmetric} (spherical) phase to the first term of the Hamiltonian and a \textit{non-symmetric} (deformed) shape to the second term. Moreover, depending on the values of $y_1$ and $y_2$ different kinds of deformation are produced. 

\section{The classical limit}
\label{sec-CL}
The study of QPTs should be strictly done in the thermodynamic limit, i.e. for an infinity number of particles. Fortunately, this kind of calculation can be easily performed through the use of the mean-field approximation which, indeed, coincides with the exact result in the large particle number limit \cite{Feng81}. The mean-field analysis of the model starts considering the product of two boson condensates, one for each fluid, 
\begin{equation}
|g\rangle=\frac{1}{\sqrt{N_1!N_2!}}(\Gamma_1^\dag)^{N_1}(\Gamma_2^\dag)^{N_2}|0\rangle, 
\end{equation} 
where $|0\rangle$ is the boson vacuum and $\Gamma_i^\dag$ the boson creation operator for the $i$ fluid defined as
\begin{equation}
\Gamma_i^\dag=\frac{1}{\sqrt{1+\beta_i^2}}~(s_i^\dag+\beta_i t^\dag_i).
\end{equation} 
The coefficients $\beta_1$ and $\beta_2$ are variational parameters associated to each fluid that, in turn, become order parameters.
  

The mean-field energy for the consistent-Q-like Hamiltonian for a
symmetric system, i.e., a system with $N_1=N_2$, in the large N
($N=N_1+N_2$) limit can be written as, 
\begin{eqnarray}
\nonumber
\frac{E(\beta_1,\beta_2,x,y_1,y_2)}{N}&=&\frac{x}{2}\left(\frac{\beta_1^2}{1+\beta_1^2}    
+\frac{\beta_2^2}{1+\beta_2^2}\right)\\
&-&\frac{1-x}{4}\left((Q_1)^2+(Q_2)^2+2 Q_1 Q_2 \right), 
\label{Egs1}
\end{eqnarray}
with
\begin{equation}
Q_i=\frac{1}{1+\beta_i^2}(2\,\beta_i+y_i\,\beta_i^2 ).
\end{equation}
Inserting these expressions for $Q_i$ in (\ref{Egs1}), we get
\begin{eqnarray}
\nonumber
\frac{E(\beta_1,\beta_2,x,y_1,y_2)}{N}&=&\frac{x}{2}\left(\frac{\beta_1^2}{1+\beta_1^2} 
+\frac{\beta_2^2}{1+\beta_2^2}\right)\\
\nonumber&-&\frac{1-x}{4}\left(\frac{1}{(1+\beta_1^2)^2}(2\,\beta_1+y_1\,\beta_1^2
  )^2 \right.\\ 
\nonumber&+&\frac{1}{(1+\beta_2^2)^2}(2\,\beta_2+y_2\,\beta_2^2)^2\\
&+&\left.2 \frac{1}{(1+\beta_1^2)}\frac{1}{(1+\beta_2^2)}
(2\,\beta_1+y_1\,\beta_1^2)(2\,\beta_2+y_2\,\beta_2^2)
  \right).
\label{energ-1}
\end{eqnarray}
Please note that, contrary to the IBM-2 case where for $\chi_\pi=-\chi_\nu$ the energy surface is invariant under the transformation $\beta_\pi \leftrightarrow \beta_\nu$, the double Lipkin model with $y_1=-y_2$ is symmetric under the interchange $\beta_1\leftrightarrow -\beta_2$. 

It could be also of interest to write down the energy for the
non-symmetric case for analyzing how the difference in the relative
number of bosons affects the mean-field energy 
\begin{eqnarray}
\nonumber
\frac{E(\beta_1,\beta_2,x,y_1,y_2)}{N}&=&x\left(F_1 \frac{\beta_1^2}{1+\beta_1^2} 
+(1-F_1) \frac{\beta_2^2}{1+\beta_2^2}\right)\\
\nonumber&-&(1-x)\left(\frac{F_1^2}{(1+\beta_1^2)^2}(2\,\beta_1+y_1\,\beta_1^2
  )^2 \right.\\ 
\nonumber&+&\frac{(1-F_1)^2}{(1+\beta_2^2)^2}(2\,\beta_2+y_2\,\beta_2^2)^2\\
&+&\left.2 \frac{F_1}{(1+\beta_1^2)}\frac{(1-F_1)}{(1+\beta_2^2)}
(2\,\beta_1+y_1\,\beta_1^2)(2\,\beta_2+y_2\,\beta_2^2)
  \right),
\label{energ-2}
\end{eqnarray}
where $F_1=\frac{N_1}{N_1+N_2}$.
Along this paper we will only consider the symmetric case.
\begin{figure}[hbt]
  \centering
\includegraphics[width=0.50\linewidth]{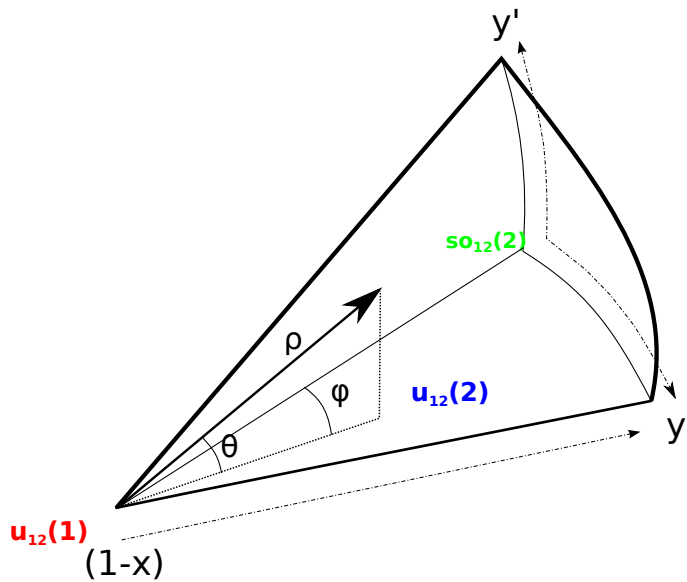}
\caption{(Color online) Representation of the two-Lipkin model
  parameter space. For completeness, the symmetries of the model are
  also indicated in the diagram.}
\label{fig-coord}
\end{figure}

\section{Numerical analysis of the phase diagram}
\label{sec-numerical}

In order to study the structure of the two-fluid Lipkin model phase diagram as a function of the control parameters, ($x, y_1, y_2$), we have found appropriate to introduce alternative control parameters, ($y, y'$), defined  by,  
\begin{equation}
y=\frac{y_1+y_2}{2},\qquad y'=\frac{y_1-y_2}{2}. 
\end{equation}
With this change, we can study the phase diagram in terms of the coordinates,
\begin{equation}
\rho=1-x;\qquad \theta=\frac{\pi}{24}(y_1-y_2)=\frac{\pi}{12}  y'; \qquad 
\phi= \frac{\pi}{24}(y_1+y_2)= \frac{\pi}{12} y.
\end{equation}  
Where we have assumed a maximum value for $|y_1|$ and $|y_2|$ equal to 2 so as the angles $\theta$ and $\phi$ are defined between $- \pi/6$ and $\pi/6$ (see Fig.~\ref{fig-coord}).

The geometric representation of the two-fluid consistent-Q-like Lipkin
Hamiltonian will be, therefore, a pyramid in this phase space. One
vertex corresponds to the $u_{12}(1)$ limit of the model ($x=1$). The
bottom plane, $y_1=y_2 \Rightarrow $ $y'=0$, corresponds to  the
$u_{12}(2)$ dynamical algebra (both fluid are combined symmetrically
into a single $u(2)$) algebra, which is equivalent at the mean-field level to
the single Lipkin model (in the IBM-2 case, the equivalent horizontal
plane represents the IBM-1, i.e., a symmetric combination of the two boson
fluids). In this plane, the line $y_1=y_2=0$ ($y =  y' = 0$) goes from
the $so_{12}(2)$ to the $u_{12}(1)$ limit. As soon as one considers
$y_1\neq y_2$ ($ y' \neq 0$) one moves in the vertical direction of
the pyramid and any present symmetry will become broken. Of special
interest is the vertical plane  $y_1=-y_2$ ($y = 0,  y' \neq 0$)
because, for this combination of parameters, 
the mean-field energy is invariant under
the transformation $\beta_1\rightarrow -\beta_2$.
Note that the two remaining vertexes do not correspond to any symmetry
of the model (in the case of IBM-2 they correspond to the $su(3)$ and $su(3)^*$ symmetries).    

As a first step to establish the phase diagram of the model, in the following of this section we will present numerical studies of different trajectories within the phase space of the model ($\rho, y, y'$) in order to identify the different phases and phase transition surfaces/lines.

To get a geometrical idea about system shapes in the different regions
of the phase space, we note that the region around the $u_{12}(1)$
vertex corresponds to values of the variational parameters $\beta_1$
and $\beta_2$ equal to zero. Since $\beta $ parameters give the weight
of the $t$ bosons in the boson condensate, $\beta =0$ implies a
condensate of spherical $s$ bosons. Consequently the phase around the
$u_{12}(1)$ vertex is called {\it symmetric} or {\it spherical}. The
corresponding spectrum will become nearly harmonic. When the system
goes well apart of the $u_{12}(1)$ vertex, both variational
parameters, $\beta_1$ and $\beta_2$ become different from zero. This
makes that the boson condensate in both fluids has a
fraction of $t$ bosons. Thus, this phase is called {\it non-symmetric}
or {\it deformed}. The horizontal plane corresponds to
$\beta_1=\beta_2$ which implies equal deformations for both fluids
what bring us back to the single Lipkin model. 
  
To study the possible phase transitions that occur in the phase
diagram we have performed numerical analysis through selected straight
trajectories in the phase space. For each of them the equilibrium
energy, the derivatives of the energy functional, and also the
equilibrium values of the variational parameters have been
analyzed. All exact calculations presented in this section correspond to
$N_1=N_2=500$ but similar studies can be done for $N_1 \neq N_2$ and
other boson numbers.
\begin{figure}[hbt]
  \centering
\begin{tabular}{cc}
    \includegraphics[width=0.4\linewidth]{path1.eps}&
  \begin{minipage}[b]{.3\textwidth}
    \includegraphics[width=\linewidth]{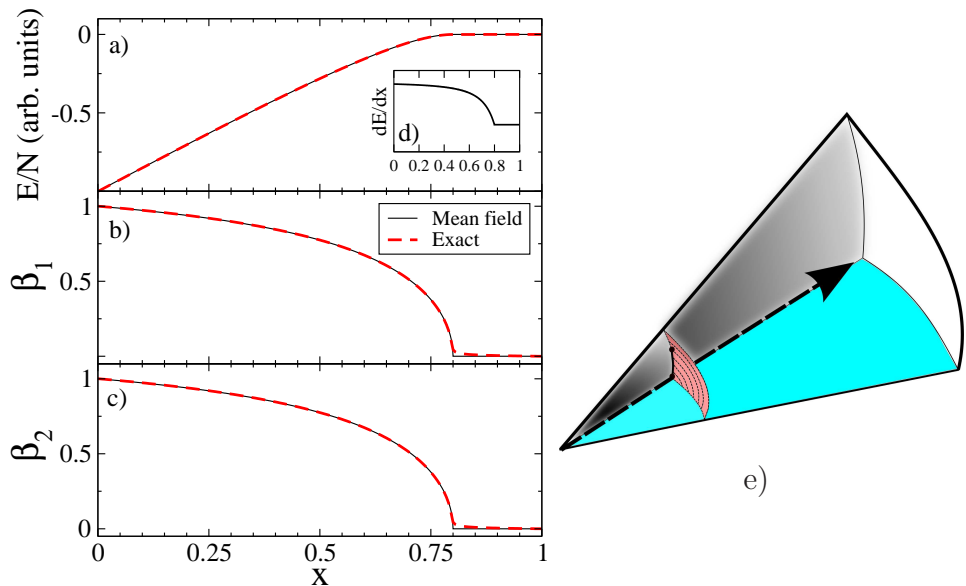}
    \vspace*{.2cm}
    \large{e)}
    \vspace*{1cm}
   \end{minipage}\\
\end{tabular}
\caption{(Color online) Transition line for $y=0$ and $y'=0$ changing
  the control parameter $x$. In the different panels we plot as a
  function of $x$: panel (a) the energy per boson (in arbitrary units),
  panel (b) the order parameter $\beta_1$ (dimensionless), panel (c)
  the order parameter $\beta_2$ (dimensionless), panel (d) $dE/dx$ in
  arbitrary units, and panel (e) representation of the trajectory in the
  control parameter space. Full thin black lines correspond to the mean-field
  results while dashed thick red lines are the exact calculation with
  $N_1=N_2=500$.} 
\label{fig-p1}
\end{figure}

\subsection{Plane $y' = 0$}

First, we start analyzing the bottom plane that corresponds to
$y_1=y_2=y$ and $y'=0$, therefore an energy surface fully equivalent
to the single Linkin case is reproduced. In Fig.~\ref{fig-p1} the line
$y_1=y_2=y=0$ and $y'=0$ is studied (see panel e), in the figure the energy per
particle (panel a), as well as, the deformation parameters (panels b and
c) as a function of the control parameter $x$ are plotted. We have also
included the function $dE/dx$ in panel (d). One can clearly see how a
phase transition setups around $x=4/5$. At the mean field level the
phase transition is established as second order since a discontinuity
appears in the second derivative of the energy (see panel (d) where
$dE/dx$ is continuous but not its derivative) and in the first
derivative of the order parameters. Note that due to symmetry
arguments $\beta_1=\beta_2$ over the whole plane. In Fig.~\ref{fig-p1}
the mean-field results (black full line) are shown together with the
exact result coming from direct diagonalization (red dashed line) for
$N_1=N_2=500$. In the exact calculation, the values of the order
parameters are extracted using this relationship,
\begin{equation}
\beta_i=\sqrt{\frac{\langle n_{t_i} \rangle}{N_i- \langle n_{t_i} \rangle}}.
\end{equation} 
Excellent agreement is found between mean-field and exact results
since the number of bosons considered in the exact diagonalization is
large enough.  

\begin{figure}[hbt]
  \centering
\begin{tabular}{cc}
    \includegraphics[width=0.4\linewidth]{path6.eps}&
  \begin{minipage}[b]{.3\textwidth}
    \includegraphics[width=\linewidth]{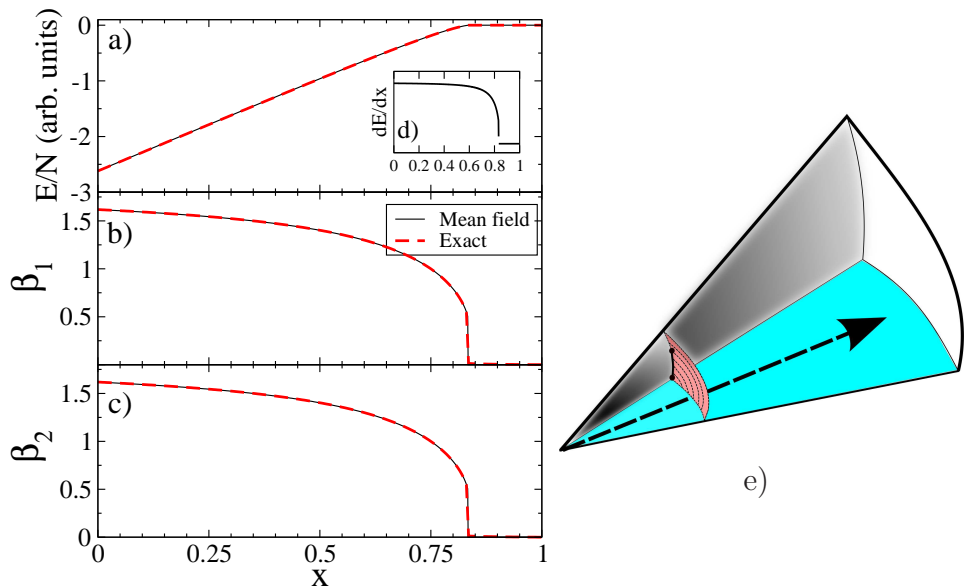}
    \vspace*{.2cm}
    \large{e)}
    \vspace*{1cm}
   \end{minipage}\\
\end{tabular}
\caption{(Color online) Same as Fig.~\ref{fig-p1} but with $y_1=y_2=y=1$ and $y'=0$.}
\label{fig-p6}
\end{figure}

In Fig.~\ref{fig-p6} we repeat the same calculation but for the line
$y_1=y_2=y=1$ and $y'=0$ (see panel e). In this case, a first order phase transition
is observed for $x_c=5/6=0.833$. The order of the phase transition is
clear from the discontinuity in the value of the order parameter at
the mentioned phase transition point, as well as for the discontinuity
in $dE/dx$ (see panel (d)). In general, for $y_1=y_2=y \neq 0$ the
phase transition is of first order and the critical point is located at
\cite{Vida06}   
\begin{equation}
x_c= \frac{4+y^2}{5+y^2}.
\label{x-crit}
\end{equation}
Consequently, the location of the critical point is shifted to
slightly larger values of $x$ as $y$ increases, while the jump of the
order parameter at the phase transition point becomes larger. 

\subsection{Volume region inside the pyramid: $y \neq 0$ and $y' \neq 0$}

\begin{figure}[hbt]
  \centering
\begin{tabular}{cc}
    \includegraphics[width=0.4\linewidth]{path8.eps}&
  \begin{minipage}[b]{.3\textwidth}
    \includegraphics[width=\linewidth]{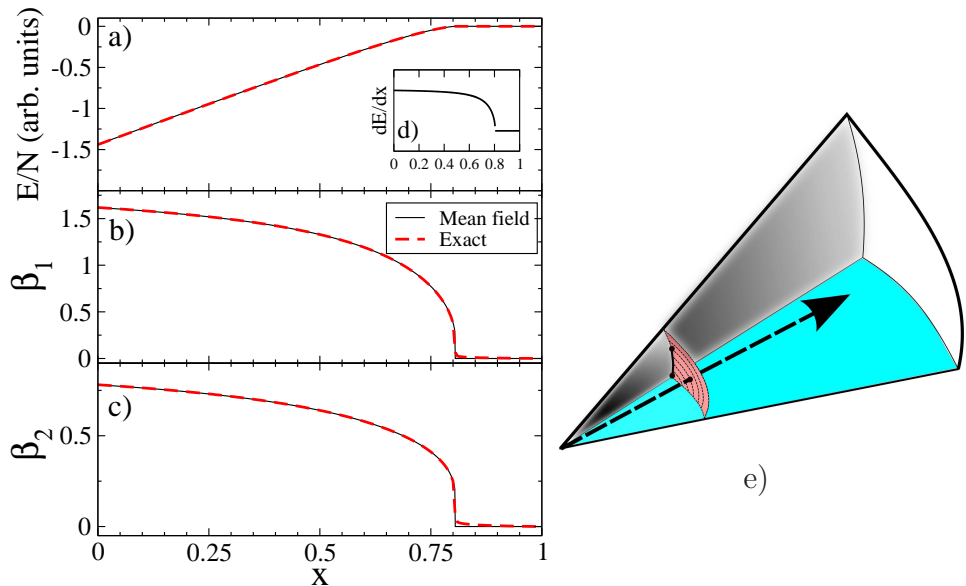}
    \vspace*{.2cm}
    \large{e)}
    \vspace*{1cm}
   \end{minipage}\\
\end{tabular}
\caption{(Color online) Same as Fig.~\ref{fig-p1} but with $y_1=1$ and $y_2=-1/2$ ($y=1/4$ and $y'=3/4$).}
\label{fig-p8}
\end{figure}

Now a trajectory going through the inner part of the pyramid is
analyzed. In particular, the case $y_1=1$ and $y_2=-1/2$, i.e.,
$y=1/4$ and $y'=3/4$, is presented  in Fig.~\ref{fig-p8} (panel e). From this
figure, it is clear that a first order phase transition is observed at
around $x_c=0.0.805$. Several trajectories inside the pyramid have been
studied with similar results (the dependence on $y$ and $y'$ of
  $x_c$ is involved and cannot be obtained in a closed form). The size of
the discontinuity depends on how far the values of $y$ and $y'$ are
from $y=0$ and $y'=0$. 
  
\subsection{The vertical plane: $y=0$}

The vertical plane corresponds to $y_1=-y_2$, what means $y=0$ and
$y'= y_1 = -y_2$ and, as we are showing below, is the most interesting
case. Several trajectories inside this plane will be presented and 
one crossing the plane from positive to negative $\beta$-values.

\begin{figure}[hbt]
  \centering
\begin{tabular}{cc}
    \includegraphics[width=0.4\linewidth]{path2.eps}&
  \begin{minipage}[b]{.3\textwidth}
    \includegraphics[width=\linewidth]{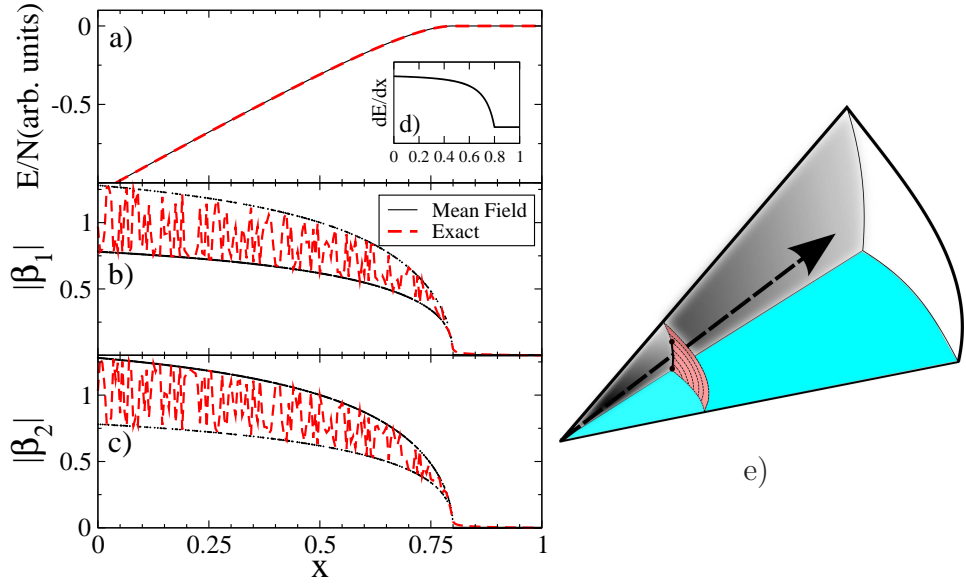}
    \vspace*{.2cm}
    \large{e)}
    \vspace*{1cm}
   \end{minipage}\\
\end{tabular}
\caption{(Color online) Same as Fig.~\ref{fig-p1} but with $y_1 = -y_2 = y' = 1/2$ and $y = 0$.}
\label{fig-p2}
\end{figure}

\begin{figure}[hbt]
  \centering
\includegraphics[width=0.70\linewidth]{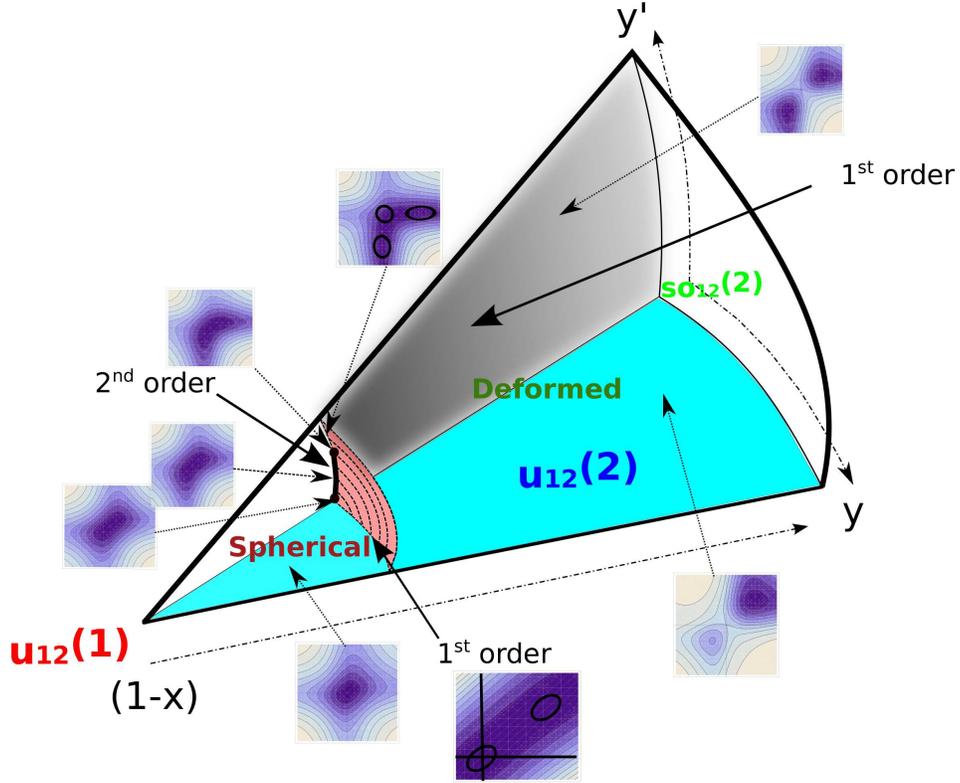}
\caption{(Color online) Phase diagram of the consistent-Q like
  two-fluid Lipkin model. In the diagram the different phases are
  represented: spherical and deformed, the first order QPT surfaces
  and the second order QPT line. Moreover, the relevant control
  parameters and dynamical symmetries also are shown.} 
\label{fig-dia}
\end{figure}

The first trajectory is the line $y_1 = -y_2 = y' = 1/2$ (panel e) and the
results are depicted in Fig.~\ref{fig-p2}. A second order phase
transition at $x_c = 0.8$ is observed. The order parameters,
coming from the exact diagonalization, show an oscillatory pattern due
to the degeneracy of two states that are related with the two minima
present in the $\beta_1-\beta_2$ plane of the mean-field energy (see
Fig.~\ref{fig-dia}). The degeneracy of two states with different deformation 
makes that the order parameter obtained from the diagonalization may
jump from one minimum to the other (between the two degenerate mean-field values). 
Indeed, for a given value of $x$ the equilibrium value of one of the
order parameters will correspond to $\beta_1=\beta_x$ and
$\beta_2=\beta'_x$, while the other to $\beta_1=-\beta'_x$ and
$\beta_2=-\beta_x$.   Note that we have taken the absolute value of
$|\beta_i|$ for a better comparison with the exact results, which are,
by definition, positive. 

\begin{figure}[hbt]
  \centering
\begin{tabular}{cc}
    \includegraphics[width=0.4\linewidth]{path3.eps}&
  \begin{minipage}[b]{.3\textwidth}
    \includegraphics[width=\linewidth]{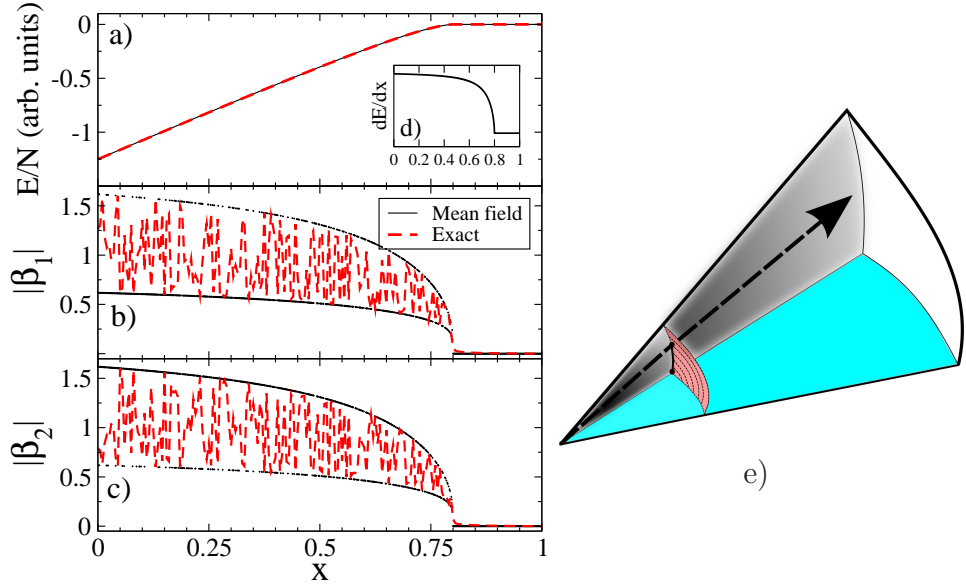}
    \vspace*{.2cm}
    \large{e)}
    \vspace*{1cm}
   \end{minipage}\\
\end{tabular}
\caption{(Color online) Same as Fig.~\ref{fig-p1} but with $y_1=-y_2=y'=1$ and $y=0$.}
\label{fig-p3}
\end{figure}

In Fig.~\ref{fig-p3} a calculation along the line $y_1 = -y_2 = y' =
1$ and $y=0$ is presented. In this case is difficult to disentangle
the order of the phase transition just looking at the energy and the
order parameters, however, in the inset panel it is clear that a
discontinuity in the second derivative of the energy exists, which, once
more, happens at around $x_c = 0.8$. In Section \ref{sec-ct-appli} we
will see in detail that, indeed, a divergence in $d^2E/dx^2$ exists
and we will try to understand the reason why there is a divergence in the
second derivative. Note that $dE/dx$ becomes vertical at $x=4/5$ from
the left side.

\begin{figure}[hbt]
  \centering
\begin{tabular}{cc}
    \includegraphics[width=0.4\linewidth]{path4.eps}&
  \begin{minipage}[b]{.3\textwidth}
    \includegraphics[width=\linewidth]{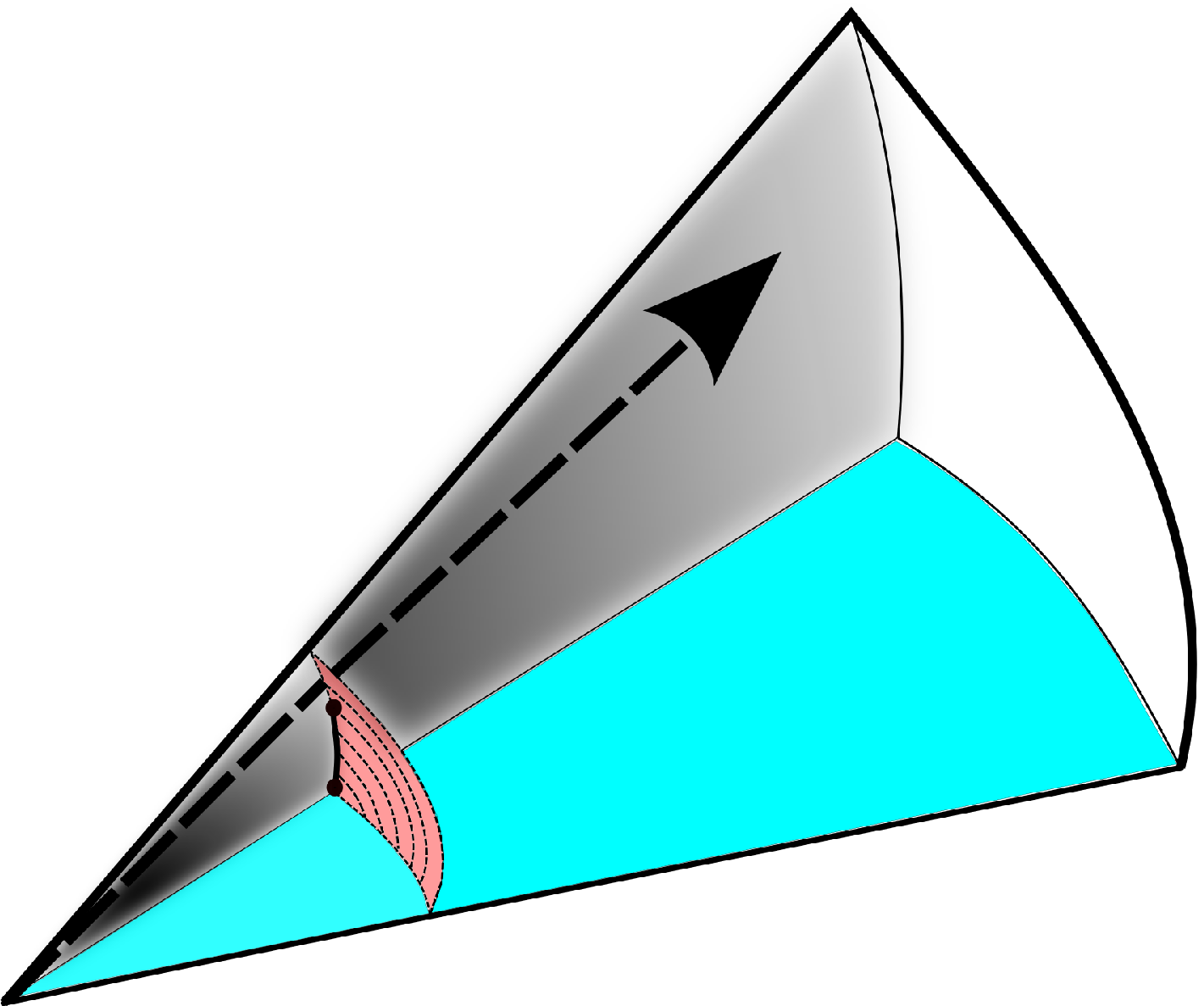}
    \vspace*{.2cm}
    \large{e)}
    \vspace*{1cm}
   \end{minipage}\\
\end{tabular}
\caption{(Color online) Same as Fig.~\ref{fig-p1} but with $y_1=-y_2=y'=3/2$ and $y=0$.}
\label{fig-p4}
\end{figure}

Finally, in Fig.~\ref{fig-p4} the calculation along
the line $y_1 = -y_2 = y' = 3/2$ and $y=0$ is shown. In this
calculation, it is easily appreciated the onset of a first order phase
transition at around $x_c = 0.81$, i.e., discontinuity in the first
derivative of the energy and in the value of the order parameters.  

After the analysis of the different paths in the plane $y = 0$,
corresponding to different $y'$ values, one is tempted to conclude
that there is a line of phase transition for values of $x$ around $x_c
= 0.8$. However,  the value $y' = 1$ separates this line into two
parts: for values $y' \le 1$ the line corresponds to a second order
phase transition, 
while for values $y' > 1$ the line is of first order. This weak conclusion, based on numerical calculations, will be confirmed in Section \ref{sec-ct-appli} through an analytic study.   

\begin{figure}[hbt]
  \centering
\begin{tabular}{cc}
    \includegraphics[width=0.4\linewidth]{path9.eps}&
  \begin{minipage}[b]{.3\textwidth}
    \includegraphics[width=\linewidth]{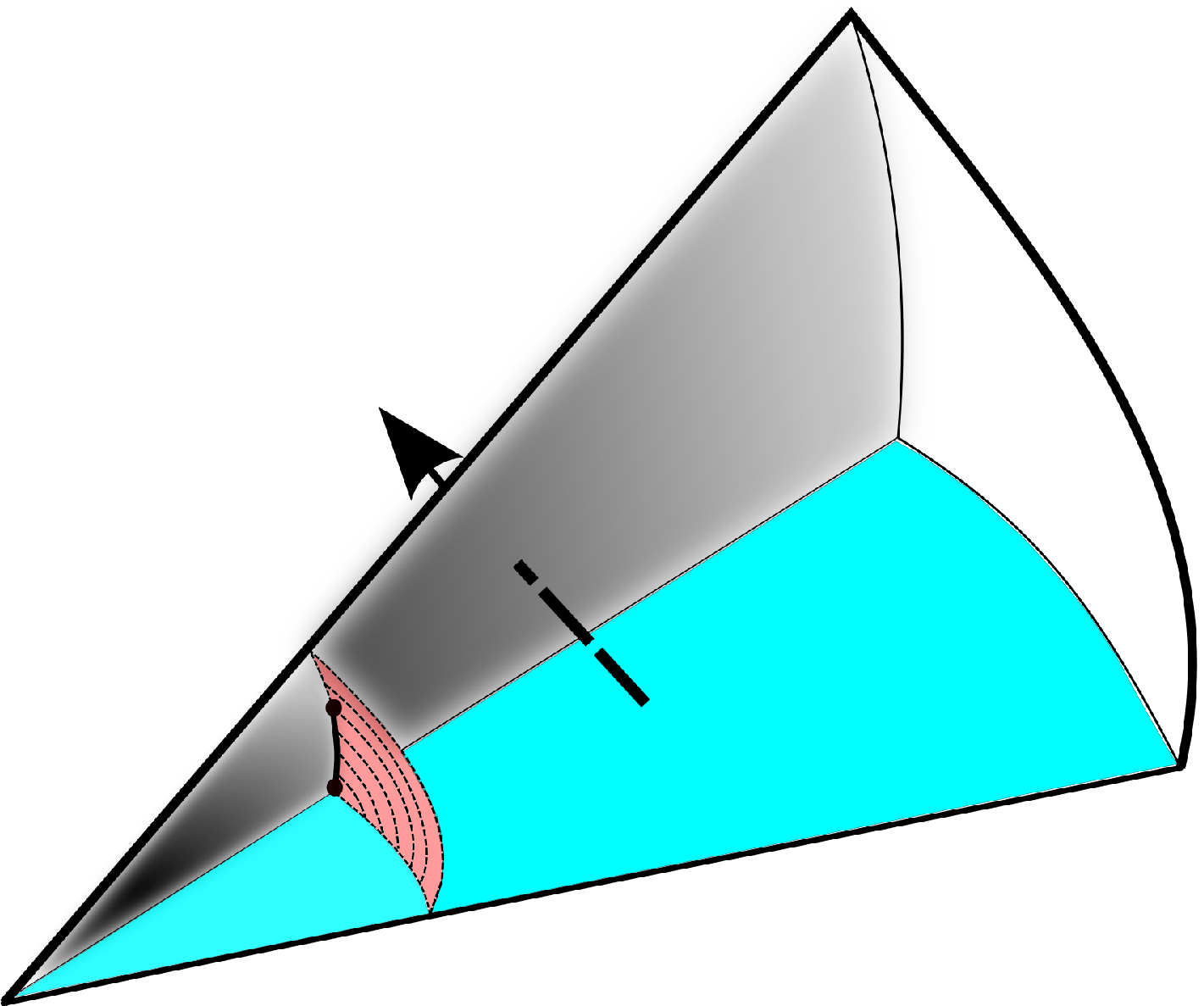}
    \vspace*{.2cm}
    \large{e)}
    \vspace*{1cm}
   \end{minipage}\\
\end{tabular}
\caption{(Color online) Transition for $x=0$ and $y_1=1$ as a function
  of $y_2$. In panel (a) the energy per boson is plotted in arbitrary
  units, in panel (b) the order parameter $\beta_1$ (dimensionless) is
  shown, in panel (c) the order parameter $\beta_2$ (dimensionless) is
  represented, in panel (d) $dE/dx$ in
  arbitrary units is plotted, and in panel (e) the representation of the trajectory in the
  control parameter space is shown. Full thin black lines correspond to the mean-field
  results while dashed thick red lines are the exact calculation with
  $N_1=N_2=500$.}
\label{fig-p9}
\end{figure}

It is worth noting that the vertical plane $y=0$, for $x<4/5$
separates two deformed regions. In order to study the transition
between both deformed regions, finally, a line crossing this vertical
surface is analyzed. Here, because of the presence of two degenerated
minima a first order phase transition for the whole vertical surface
in the deformed phase is expected. This is fully confirmed in
Fig.~\ref{fig-p9} (parameters $x=0.5$, $y=1$), where the first order
phase transition appears for $y_2 = -1$. Please, note that in
Fig.~\ref{fig-p9} we have changed by hand the value of $\beta$ coming
from the exact calculation (it is, by definition, always positive)
for a better comparison with the mean-field results.    

Combining all the preceding evidences one gets the phase diagram
depicted in Fig.~\ref{fig-dia}, where one can appreciate a first order
phase transition surface separating the symmetric (spherical) and non-symmetric
(deformed) phases and the first order phase transition vertical
surface separating 
two different deformed phases. We will see in next section that the
intersection line between both surfaces, from $y'=0$ up to $y'=1$ is a
second order phase transition line, while for larger values of $y'$ it
becomes first order. Note that the phase diagram can be extended to
negative values of $y$ and $y'$, with the first order phase transition
surface separating {\it spherical} and {\it deformed} phase extended
to four quadrants and the vertical first order phase transition
surface and the second order phase transition line extended to
negative values of $y'$.  
  
\section{Local Taylor expansion and catastrophe theory}
\label{sec-taylor}

\subsection{Fundamentals}
\label{sec-ct-fund}
Once the main structure of the phase diagram of the model is known numerically, it is necessary to perform an analytic study to determine unambiguously the order of the phase transitions of surfaces and lines appearing in the phase diagram and to understand why the QPT areas are precisely located there. To carry out this task we will make use of catastrophe theory (CT) \cite{Thom75}, which is an ideal tool for such an end.

In general, the aim of CT is to study a given potential, $V(\vec{x}, \vec{\lambda})\in\Re$ (in our case the mean-field energy surface of the model) or a family of potentials that are function of a set of order parameters, $\vec{x} \in\Re^n$  and  that depend on a set of control parameters, $\vec{\lambda}\in\Re^r$, and to study the qualitative behavior of the potential, e.g., number of minima and maxima, as a function of the control parameters. To proceed one should start looking for the stationary points (also known as  critical points), i.e., those whose gradient vanish and classify them according to their stability: i) points where the determinant of the Hessian matrix  is different from zero, called isolated, non-degenerated or Morse points, and ii) points where the determinant of the Hessian matrix is zero, called non-isolated, degenerated or non-Morse points. In summary, points of a family of smooth potentials can be classified according to their gradient and Hessian matrix ${\cal H}$ as:
\begin{itemize}
\item Regular points: $\nabla V\neq 0$.
\item Morse points (isolated critical points):
$\nabla V= 0$ and $|{\cal H}|\neq 0$.
\item Non-Morse points (degenerated critical points):
$\nabla V= 0$ and $|{\cal H}|= 0$.
\end{itemize}

Morse theorem \cite{Post78,Gilm81} guarantees that around a Morse point, a smooth potential is equivalent to a quadratic form, performing a smooth non-linear change of variables. Therefore, the potential is stable under small perturbations around Morse points. At non-Morse points the potential cannot be written as a quadratic form because the Hessian matrix has at least one zero eigenvalue. Around non-Morse points CT will provide useful information on how the qualitative shape of the potential will evolve under small variations of the order parameters.

In the case of several order parameters, Thom's {\it splitting lemma} \cite{Thom75} guarantees that a smooth potential at non-Morse points can be written as a sum of a quadratic form, associated to the subspace with nonzero eigenvalues, plus a function containing the variables associated to the zero eigenvalues of the Hessian matrix. 

The first step in the CT program is to find out the critical points of the energy surface ($\nabla E=0$). Among them, the most important is the most degenerate one, i.e., the point where most successive derivatives vanish. This point is the fundamental root taking place at definite values of the control parameters which we will call critical values.  We next proceed making use of a Taylor expansion of the energy surface around the fundamental root. A Taylor expansion around such a point is also valid for the critical points that {\it arise} from the fundamental root when the degeneracy is broken. Depending on the degeneracy of the fundamental root the number of extremes that can be analyzed simultaneously will change. It is worth to note that the different minima related with the appearance of a critical phenomenon arise from a degenerated non-Morse point.

When the potential depends on several  variables, as the case for the two-fluid Lipkin model is,  it is important to separate the variables into two sets, depending on how Hessian eigenvalues behave. On one hand, one has the variables associated to the subspace with vanishing Hessian eigenvalues, called  {\it bad} or {\it essential} variables, while, on the other hand, there is a set of variables related to the non-vanishing Hessian eigenvalues, called {\it good} or {\it non-essential} variables. Therefore, as a consequence of the {\it splitting lemma} \cite{Thom75}, the potential could be separated into a part depending on the {\it essential} variables and into another part depending on the {\it non-essential} ones by rewriting it in terms of the eigenvectors of the Hessian matrix \cite{Gilm81}. The appearance of  critical phenomena will be associated exclusively with the behavior of the {\it essential} variables, i.e., the variables that can be identified as order parameters of the system.

\subsection{Application to the two-fluid Lipkin model}
\label{sec-ct-appli}
According to Eqs.~(\ref{energ-1}) and (\ref{energ-2}) the most degenerated critical point for the two-fluid Lipkin model (\ref{Hcqf}) corresponds to $\beta_1=0$ and $\beta_2=0$ (all derivatives up to fourth order vanish for an appropriated set of parameters) and, taking into account the shape of the phase diagram, all the critical points that can eventually arise in the energy surface are born from this most degenerated critical point. Since there are two shape variables it is necessary to construct, first, the Hessian matrix associated to the energy surface (\ref{energ-1}), 
\begin{equation}
\label{hess1}
\cal{H}=
\left (
\begin{array}{cc}
\partial^2E/\partial\beta_{1}^2
& \partial^2E/\partial\beta_{1}\partial\beta_{2}\\
\partial^2E/\partial\beta_2\partial\beta_1 & \partial^2E/\partial\beta_2^2
\end{array}
\right )=
\left (
\begin{array}{cc}
3x-2 & 2x-2\\
2x-2 & 3x -2
\end{array}
\right ).
\end{equation}
The two eigenvalues are $5x-4$ and $x$, and the corresponding eigenvectors are,
\begin{eqnarray}
\beta_a&=& \frac{1}{2}~ (\beta_1+\beta_2),\\
\beta_b&=& \frac{1}{2}~( \beta_1-\beta_2).
\end{eqnarray}
The eigenvalue associated to $\beta_a$ vanishes for $x=4/5$ while the
one associated to $\beta_b$ only vanishes for the trivial case
$x=0$. Therefore, the {\it essential} variable turns out to be
$\beta_a$, while $\beta_b$ becomes the {\it non-essential} one, i.e., the origin in this variable behaves as a Morse point.

Next step is to carry out a Taylor expansion in $\beta_a$ and $\beta_b$ around zero. Because we use $\beta_a$ and $\beta_b$, the quadratic term $\beta_a\beta_b$ will not be present in the Taylor expansion. Note that we are considering the case $N_1=N_2$.
\begin{eqnarray}
\nonumber
\frac{E(x,y,y',\beta_a,\beta_b)}{N}&=& (5 x-4) \beta_a^2 +4(x - 1)y\beta_a^3 +
\left(8-9 x+y^2(x-1)\right)\beta_a^4 
+ \Theta(\beta_a^5) +x\beta_b^2 +\Theta(\beta_a\beta_b^2, \beta_b\beta_a^2) .
\end{eqnarray}
In order to cancel the higher order terms ($\beta_a^i \beta_b^j$ with $j > 1$) we have to implement a nonlinear transformation in $\beta_b$ 
\begin{equation}
\tilde{\beta_b}=\beta_b+\sum_{i+j>1}a_{ij}\beta_a^i \beta_b^j .
\end{equation}
After imposing the cancellation of the crossing terms and determining
the value of $a_{ij}$, we get the next expression that is valid in the
neighborhood of $\beta_a=0$, but for any value of $\beta_b$ (note that in order to simplify the notation we continue referring to the {\it non-essential} variable as $\beta_b$ instead of $\tilde{\beta_b}$), 
\begin{eqnarray}
\nonumber
\frac{E(x,y,y',\beta_a,\beta_b)}{N}&=& (5 x-4)\beta_a^2 +
 4 (x-1) y\beta_a^3
+
\left(8-9 x+y^2(x-1)-\frac{16 {y'}^2}{x}(x-1)^2
\right)\beta_a^4 \\
\nonumber
&+&
 \frac{8(x-1) y \left((6 {y'}^2-1) x^2
-14 {y'}^2 x+8 {y'}^2\right)}{x^2}\beta_a^5\\
\nonumber
&+&
 \frac{1}{x^3}\left((64 {y'}^4-384 {y'}^2-2 y^2
(82{y'}^2+1)+13) x^4+2(-96 {y'}^4+576{y'}^2\right.\\ 
\nonumber
&+&
y^2 (372 {y'}^2+1)-6) x^3+
4 {y'}^2 (48 ({y'}^2-6)-313 y^2)x^2\\
\nonumber
&+&
\left. 32 {y'}^2 (29 y^2-2 {y'}^2+12) x-256 y^2{y'}^2\right)\beta_a^6\\
&+& O(\beta_a^7)+x\beta_b^2 .
\label{taylor1}
\end{eqnarray}

The number of lower order terms that are kept in the Taylor expansion
without loosing substantial information with respect to the original
function (problem of {\it determinacy}) is determined studying the
terms in the Taylor expansion that can be canceled out with
appropriated particular values of the control parameters. For
Eq.~(\ref{taylor1}) the values of the control parameters,  $x=4/5$, $y=0$  and
$y'=1$ cancel all terms up to $\beta_a^5$. Therefore, the dominant
remaining term is $\beta_a^6$ and it is said that the function is
$6-determined$. Hence, the number of {\it essential} parameters will
be $3$. Consequently, the relevant elementary catastrophe of this
model is the {\it butterfly} (A$_{+5}$) \cite{Gilm81}. It is worth
mentioning that the {\it butterfly} has a {\it codimension} equal to
$4$, i.e., the number of essential parameters is $4$. In our case, due
to the function symmetry, the number of parameters is only $3$. In
general, depending on the values selected for the control parameters,
the potential energy may have three, two, or one local minima. Please
note that Eq.~(\ref{taylor1}) does not correspond to the canonical
form of the {\it butterfly} because it presents a fifth order term
instead of the first order one. However, one always can perform a {\it
  shift} transformation in the $\beta_a$ variable to recover the canonical form. 

In order to determine the order of the phase transitions, already
studied numerically in the preceding section, we can take advantage of
Eq.~(\ref{taylor1}). In general, for any situation with $y \neq 0$ the
cubic (and fifth) term always survives for any value of $y'$ and, therefore, the
phase transition will become first order. The reason is simple, the
presence of a cubic (and fifth) term guaranties the possible
appearance of several critical points (i.e., a region of coexistence),
three critical points (two minima and one maximum) when the
$\beta_a^2$ coefficient and 
the $\beta_a^4$ coefficient are positive and five critical points (three
minima and two maxima) when the
$\beta_a^2$ coefficient is positive and the $\beta_a^4$ coefficient
is negative (see below for more details). Note that in our case, the
$\beta_a^6$ coefficient is always positive.
Indeed, this
particular situation is precisely the one necessary to develop a first
order phase transition. This happens in almost the whole surface separating the
{\it symmetric} (spherical) and {\it non-symmetric} (deformed) phases.

To know the character of the vertical surface, $y=0$ (with $x<4/5$),
one can note that the lowest leading terms are: $\beta_a^2$  with
negative coefficient and  $\beta_a^4$ with positive coefficient. This
potential gives rise to two degenerated minima symmetric with respect
to the origin, $\beta_a=0$. As soon as one perturbs the system,
changing $y$ (to either positive or negative values) the degeneracy is
broken and one of the deformed minima is lower in energy. Therefore,
this situation corresponds to a first order phase transition since the order parameter will jump suddenly from one to the other minimum.

It is also of interest to see how one can recover the case
corresponding to the single Lipkin, i.e., $y'=0$. For this case
(horizontal plane) the energy surface reads as, 
\begin{eqnarray}
\frac{E(x,y,y',\beta_a,\beta_b)}{N}&=&(5 x-4)\beta_a^2  +
 4(x-1) y\beta_a^3
+
\left(y^2(x-1)+8 -9x\right)\beta_a^4 
+ O(\beta_a^5)+x\beta_b^2 ,
\end{eqnarray}
where one can easily single out that for the line $x=4/5$, values $y
\neq 0$ produce a first order phase transition since the cubic term is
present. For 
the particular value $y=0$ the cubic term vanishes and, therefore, the
transition is no longer of first order, but of second order.
 
Finally, a most interesting case is the intersection line between the
surfaces $y = 0$ (vertical plane, separating two regions of different
deformation) and $x \approx 4/5$ (spherical surface separating
spherical from deformed shapes). For this situation, the energy
functional is written as, 
\begin{eqnarray}
\nonumber
\frac{E(x,y',\beta_a,\beta_b)}{N}&=& (5x-4)\beta_a^2 +
\left (8-9x -\frac{16(x-1)^2y'^2}{x}\right)\beta_a^4\\
&+&
\frac{1}{2}\left(26x-24+\frac{128(x-1)^3 y'^2(y'^2-6)}{x^2}\right)\beta_a^6 
+
 O(\beta_a^7)+ x\beta_b^2 .
\label{lasexta}
\end{eqnarray}
\begin{figure}[hbt]
  \centering
\includegraphics[width=0.50\linewidth]{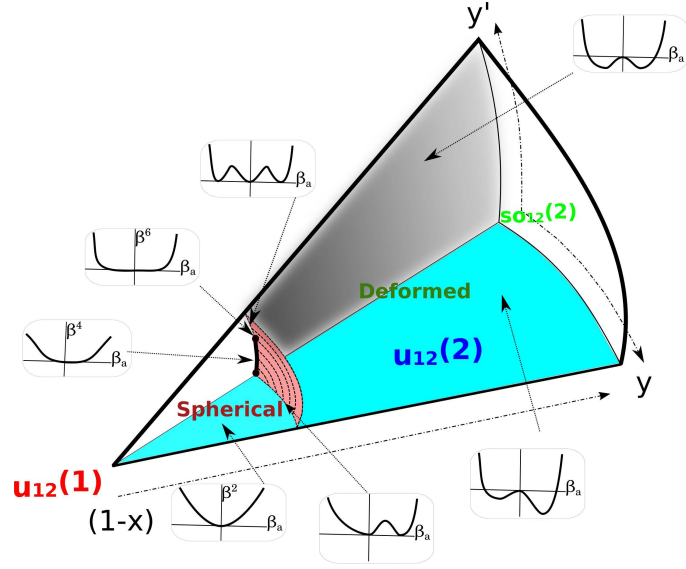}
\caption{(Color online) Same as Fig.~\ref{fig-dia} but including the
corresponding energy curves as a function of the {\it essential} order
parameter $\beta_a$.}
\label{fig-dia-pepe}
\end{figure}
This situation looks like the case $y = y' = 0$, because no odd terms
appear in the expansion and only the onset of a second order phase
transition is expected. However, there are fundamental
differences. The key point to disentangle the stability structure of
the energy surface is the sign of the fourth order coefficient. The
phase transition  at $x=4/5$ is indeed of second order if the fourth order
coefficient remains positive but will change to first order
otherwise. There is a critical value for $y'$ for which the fourth
order coefficient vanishes, i.e., $y'=1$ and $x=4/5$. At this point,
the only term that survives in the energy functional is the sixth
order term. The most important consequence is that at this point the
energy surface is very flat (as $\beta_a^6$). Going to values with $y'>1$,
the fourth order coefficient changes to negative sign, which implies
that there is no longer a second order phase transition, but a first
order one since, in this case, there is a sudden change in the order
parameter when crossing the QPT point. To understand this fact let us
write in a more compact form the Taylor expansion (\ref{lasexta}) as, 
\begin{equation}
E=\frac{A}{2} \beta^2+\frac{B}{4} \beta^4+\frac{C}{6} \beta^6, 
\label{eq-butterfly}
\end{equation}
where $C>0$.  According to Eq.~(\ref{lasexta}): $A>0$ for $x>4/5$,
$A<0$ for $x<4/5$ and $A=0$ for $x=4/5$; at $x=4/5$, $B>0$ for $y'<1$,
$B<0$ for $y'>1$, and $B=0$ for $y'=1$; for  $x\approx 4/5$ $C$ is
always positive. Equation $dE/d\beta=0$ has as solutions,
\begin{eqnarray}
\beta&=&0 , \nonumber \\
\beta^2&=& \frac{-B \pm \sqrt{B^2-4 A C}}{2C}.
\end{eqnarray}

The spherical solution ($\beta = 0$) corresponds to a minimum if $A >
0$, i.e., $x>4/5$ (to a maximum if $A < 0$,  i.e., $x<4/5$ ),
irrespective of the $B$ sign, i.e., independently of the $y'$
value. For  $A < 0$ and $B > 0$ (which occurs for $|y'| < 1$) two
deformed critical points exist (symmetric with respect to the origin),
which correspond to minima since, as discussed above, $\beta=0$
corresponds, in this case, to a maximum. Note that for $A = 0$ the two
deformed minima merge into a flat spherical one, never
coexisting several minima. Therefore, the line $y = 0$ and $A = 0$
corresponds to a 
second order phase transition while $|y'| < 1$. 

For $A \gtrsim 0$ and $B < 0$ (which occurs for $y' > 1$ and
$x\gtrsim 4/5$) five critical points coexist for $B^2 > 4 AC$, one
corresponds to the spherical minimum, and other two correspond to two
deformed minima (symmetric with respect to the origin,
$\beta_a=0$). The other two extremes correspond to maxima (symmetric
with respect to the origin). The particular region $y = 0$, $A = 0$
($x=4/5$) for values  $y' > 1$, is a region of coexistence of three
minima, one spherical and two deformed. At the critical point all
three minima are degenerated. As a consequence, a first order phase transition develops around this line. The first order phase transition line is defined then by
\begin{equation}
A=\frac{ B^2}{4 C},
\end{equation} 
and is bounded by the spinodal ($(\partial^2 E/\partial
\beta^2)_{\beta=0}=0$) and the antispinodal ($(\partial^2 E/\partial \beta^2)_{\beta=\beta_c\neq 0}=0$) lines given by,
\begin{eqnarray}
A&=&0 ~~~~~~~~~~~~~~ \textrm{(spinodal)},\\
A&=&\frac{3 B^2}{16 C} ~~~~ \textrm{(antispinodal)}.
\end{eqnarray}
For the case $B^2 < 4 A C$ only the spherical minimum exists.

It is worth analyzing what happens at the special line $y = 0$ and $y'=1$. For this line $B = 0$ and the energy surface presents three critical points when $A < 0$ ($x < 4/5$),
\begin{eqnarray}
\beta&=&0,\\
\beta&=&\pm(\frac{-A}{C})^{1/4}.
\label{betasexta}
\end{eqnarray}      
The first one, spherical, corresponds to a maximum and the second and
the third, deformed and symmetric with respect to the origin, correspond to minima. For $A > 0$ ($x > 4/5$) only the minimum at $\beta=0$ survives. In order to study the QPT at $x = 4/5$, one can write the energy at the equilibrium $\beta-$value (\ref{betasexta}) that is,
\begin{eqnarray}
E&=&\frac{(-A)^{3/2}}{3 \sqrt{C}} ~~~~~~~~~~~~~~~~~~ \textrm{for} ~~ A<0 , \\
E&=&0 ~~~~~~~~~~~~~~~~~~~~~~~~~~~~~\textrm{for} ~~ A>0 .
\end{eqnarray}  
Its first derivative with respect to $A$ is
\begin{eqnarray}
\frac{d E}{d A}&=&\frac{-\sqrt{-A}}{2\sqrt{C}}~~~~~~~~~~~~~~~ \textrm{for} ~~ A<0 , \\
\frac{d E}{d A}&=&0 ~~~~~~~~~~~~~~~~~~~~~~~~~\textrm{for} ~~ A>0 .
\end{eqnarray} 
Therefore, the transition is not of first order. Performing the second order derivative,
\begin{eqnarray}
\frac{d^2 E}{d A^2}&=&\frac{-1}{4\sqrt{-A C}},~~~~~~~~~~~~~~~~~~~ \textrm{for} ~~ A<0 , \\
\frac{d^2 E}{d A^2}&=&0 ~~~~~~~~~~~~~~~~~~~~~~~~~~~~~~~\textrm{for} ~~ A>0 .
\end{eqnarray} 
Therefore, a discontinuity appears in the second derivative with respect to the control parameter. Indeed in the deformed side the second derivative diverges to $-\infty$. 

\begin{figure}[hbt]
  \centering
\includegraphics[width=0.50\linewidth]{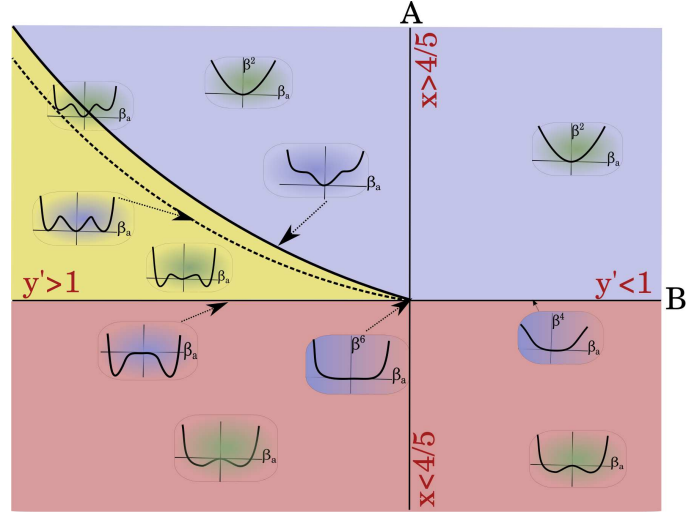}
\caption{(Color online) Representation of the {\it butterfly} catastrophe
(without odd-power terms), Eq.~(\ref{eq-butterfly}), as a function of
the control parameters $A$ 
and $B$. The energy curves that characterize every region are depicted,
besides, the curves corresponding to lines/points with degenerated
critical points are also depicted.}
\label{fig-butterfly}
\end{figure}

In Fig.~\ref{fig-dia-pepe} we show, once more, the phase diagram, but
in this case plotting the corresponding energy curves as a function of
the value of the {\it essential} variable. In this figure one can
appreciate in a cleaner way how the first order vertical plane is
related to two symmetric degenerated minima, symmetric with respect to
$\beta_a=0$, while the first order surface separating spherical and
deformed phases corresponds also to two degenerated minima, spherical
and deformed. The line $x=4/5$, $y=0$, $y'<1$ corresponds to a
$\beta_1^4$ energy curve, i.e., to a {\it cusp} line. The point
$x=4/5$, $y=0$, $y'=1$ corresponds to a 
$\beta_1^6$ energy curve and, finally, a first order phase transition
line appears for
$x\approx 4/5$, $y=0$, $y'>1$ with three degenerated minima. The
behavior of this first order phase transition area is explained in detail in
Fig.~\ref{fig-butterfly}, where as a function of the control parameters
$A$ and $B$ is depicted the phase diagram, separating the areas
corresponding to spherical (blue area), deformed shapes (red
area) or coexistence area (yellow area). Also spinodal, antispinodal
and first order line are shown.   

At the point $A=0$, $B=0$, i.e., $x=4/5$, $y=0$, and $y'=1$ the second
order phase transition line ($B>0$) and the spinodal, antispinodal and first order
phase transition merge. This point is known as tricritical point while the first
order phase transition line corresponds to a triple point curve where three minima are
degenerated and coexist. 

The inclusion of a third (and a fifth) order term in the potential
(\ref{eq-butterfly}) will break the symmetry of the function. The
consequence will be the appearance of a coexistence region for the
{\it cusp} line and, therefore its transformation in a first order
phase transition line. In the case of the coexistence area with $B<0$
the asymmetry generated in the energy curves will make impossible the
degeneracy of three minima, but however, will continue being possible the
degeneracy of the spherical and one of the deformed minima. 
Moreover the spinodal line is still at $A=0$ too, though the antispinodal
one will be shifted. As a consecuence, in this case, the phase
transition is still of first order.

\section{Summary and conclusions}
\label{sec-conclusions}
In this work the mean-field energy surface of the consistent-Q like
double Lipkin Hamiltonian has been studied. The analyzed Hamiltonian
resembles the IBM-2 Hamiltonian of interest in Nuclear Physics. The
phase diagram of the model has been established both numerical and
analytically. The mean-field numerical calculations have been compared
with direct diagonalizations and good agreement has been reached. The
analytical study has been performed using the catastrophe theory and
it has been found that the energy can be successfully described by the
{\it butterfly catastrophe}. 

Therefore, the phase diagram of the model has been obtained, including
phases, locations of the QPT phase transitions and their orders. In
particular, there are three phases: spherical and two different
deformed ones. The surface separating  {\it spherical} and both {\it
  deformed} phases is of first order: two minima, one spherical and one
deformed, which are degenerated at the phase transition
point. Moreover, the vertical plane separating both deformed phases is
also of first order: two deformed minima with different deformations are
degenerated and separated by a maximum at $\beta_a=0$. These two
surfaces intersect in the line ($x=4/5$, $y=0$), this is of second order
for $0 <y' \le 1$ and transforms to a first order phase transition for $y'>1$.  The
part of the line ($x=4/5$, $y=0$, $0 <y' < 1$) corresponds to a flat
surface (goes as $\beta_a^4$). At the point ($x=4/5$, $y=0$,
$y'=1$) the energy surface is even flatter, goes as
$\beta_a^6$. Finally, in the part of the line ($x\approx 4/5$, $y=0$, $y' >
1$) three degenerated minima coexist (one spherical and two deformed
ones). The point $x=4/5$, $y=0$ and $y'=0$ corresponds to a tricritical
point in the language of the Ginzburg-Landau theory for phase transitions.
 
\section{Acknowledgment}
This work has been supported by the Spanish Ministerio de
Econom\'{\i}a y Competitividad and the European regional development
fund (FEDER) under Project No. FIS2011-28738-C02-01,
FIS2011-28738-C02-02, FIS2014-53448-C2-1-P, FIS2014-53448-C2-2-P, and
by Spanish Consolider-Ingenio 2010 (CPANCSD2007-00042).

\end{document}